\documentclass[letter,twocolumn]{jpsj3} 
\usepackage{graphicx}
\usepackage{color}
\def\i{i}
\def\vec#1{\mbox{\boldmath $#1$}}
\def\ket#1{\left|#1\right\rangle }
\def\bra#1{\left\langle #1\right|}
\def\braket#1#2{\langle #1|#2\rangle}
\def\d#1{\,{d}#1}
\def\e#1{\,{e}#1}
\def\refeq#1{eq.~(\ref{#1})}

\title{Fractionally quantized Berry phases of magnetization plateaux in spin-$1/2$ Heisenberg multimer chains}
\author{Isao \textsc{Maruyama}$^1$ and Shin \textsc{Miyahara}$^2$}

\inst{$^1$Department of Information and Systems Engineering, Fukuoka Institute of Technology, 3-30-1 Wajiro-higashi, Higashi-ku, Fukuoka 811-0295, Japan \\
$^2$Department of Applied Physics, Fukuoka University, 8-19-1, Nanakuma, Jonan-ku, Fukuoka 814-0180, Japan
}
\abst{
We study the fractionally quantized $Z_N$ Berry phase, 
$\gamma_N=0, {2\pi \over N}, {4\pi \over N},\ldots, {2(N-1)\pi \over N}$,
to characterize local $N$-mer spin structures at magnetization plateaux
in spin-1/2 Heisenberg 
multimer ($N$-mer) models, i.e.,  highly frustrated $N$-leg ladder models,
which are generalizations of an orthogonal dimer chain and 
have exact ground states in the strong multimer coupling region.
We demonstrate that all $N$ types of Berry phases, which characterize magnetization-plateau phases, appear in a magnetic phase diagram when $N=2$ and $4$.
We show that magnetization plateau with magnetization $\langle m\rangle$ and $D$-fold degenerated states
has $\gamma_N=\pi (\langle m\rangle-1) D$,
except for the Haldane phase with $\gamma_N=0$.
In addition, we find that a complementary $Z_N$ Berry phase becomes non-zero in the $S=N/2$ Haldane phase for $N=2$ and 4.
Because the exact quantization of the $Z_N$ Berry phases
is protected by the translational (or rotational) symmetry along the rung direction,
the $Z_N$ Berry phase has the potential to be applied for a wide class of magnetization plateaux in coupled multimer systems.
}

\begin{document}
\maketitle


Although symmetry breaking has been successfully described by local
order parameters, topological phases cannot be characterized by local
order parameters.  One of the historical examples is the hidden $Z_2\times
Z_2$ symmetry breaking of the Haldane phase\cite{PRB.45.304},
which requires new quantities for characterization, such as
a non-local string order parameter\cite{PRB.40.4709},
twisted-order parameter\cite{PRL.89.077204},
entanglement spectrum\cite{PRB.81.064439}, 
and a quantized Berry phase\cite{JPSJ.75.123601}.

The quantized Berry phase defined through local flux 
is one of the new topological invariants. 
A historical example of topological invariants is the Chern number for integer quantum Hall effects\cite{PRL.49.405}.
As an advantage, 
various types of Berry phase can be defined depending on the choice of the local flux 
and, thus, the quantized Berry phase has the potential
to characterize various gapped phases. 
In fact, 
the Berry phase has more variety than anti-periodic boundary conditions,
as shown in a recent study using the $Z_2$ Berry phase\cite{PRB.94.205112}.
In addition, fractional $Z_N$ quantization of the Berry phase\cite{EPL.95.20003
,AX.1508.00960,AX.1709.01546,AX.1806.10767} 
can characterize $N$-types of phases.
However, it has not been demonstrated that 
$N$ different
phases, which are characterized by a $Z_N$ Berry phase defined 
in the same location, exist in a magnetic phase diagram.

In this letter, we show that fractionally quantized $Z_N$ Berry phases
successfully characterize magnetization-plateau phases including the Haldane phase in  magnetic phase diagrams of
spin-$1/2$ Heisenberg 
$N=2$ and $4$ multimer chains, that is, highly frustrated $N$-leg ladders\cite{EPJB.15.227}.
In a strong multimer coupling region of the model, the ground state is exactly obtained 
as a direct product state. 
Because the ground states are rigorously written, the 
model Hamiltonian is a useful reference to understand 
the role of the $Z_N$ Berry phase as well as
the physics of magnetization, 
magnetization plateaux, and jumps.
The $N=2$ model is equivalent to the edge-shared tetrahedral chain\cite{PRB.43.8644},
which is a one-dimensional version of the Shastry-Sutherland model\cite{PB.108.1069}
in a class of the orthogonal dimer models\cite{JPCM.15.327}.
Because the $Z_2$ Berry phase can detect the dimer singlet
in the two-dimensional orthogonal dimer model,\cite{JPCONF.320.012019}
we introduce the same $Z_2$ Berry phase in the dimer ($N=2$) chain under the external magnetic fields.
On the other hand, there has been no previous study on the $Z_N$ Berry phase for multimer ($N$-mer) chains 
for the $N \neq 2$ cases. Here, we mainly study the $Z_4$ Berry phase for the tetramer ($N=4$) chain 
as a typical example of the $N \neq 2$ cases.


Let us define the 
isolated spin-$1/2$ multimer ($N$-mer) 
Hamiltonian at the position $x$ along the leg:
\begin{eqnarray}
  h_x = \left\{
    \begin{array}{lc}
      J \vec{s}_{x,1}\cdot \vec{s}_{x,2}, & \mbox{for $N=2$}
      \\
     \displaystyle 
      J \sum_{y=1}^N \vec{s}_{x,y}\cdot \vec{s}_{x,y+1} , & \mbox{for $N>2$}
    \end{array}
    \right.
    .
\end{eqnarray}
We adopted the periodic boundary condition (PBC), $\vec{s}_{x,y}=\vec{s}_{x,y+N}$.
Note that we use a one-dimensional ring structure for an $N$-mer for simplicity,
which is slightly different from the definition in Ref.~\citen{EPJB.15.227} for $N \geqq 4$.
By using the
total spin of each local multimer $\displaystyle \vec{S}_x  = \sum_{y=1}^N\vec{s}_{x,y}$, 
the total Hamiltonian ${\cal H}$ of the $N$-mer with system size $L$
under the magnetic field $H$
is defined as
\begin{eqnarray}
  {\cal H}=\sum_{x=1}^{L} \left( h_{x}  + J' \vec{S}_x \cdot \vec{S}_{x+1}  - H S^z_{x}\right).
  \label{eq:H}
\end{eqnarray}
Here, we adopted the PBC $\vec{s}_{x,y}=\vec{s}_{x+L,y}$ along the legs.
We restricted ourselves to the case $J > 0$ and $J' > 0$ and used $J$ as the unit of energy.
\begin{figure}
\includegraphics[width=0.5\textwidth]{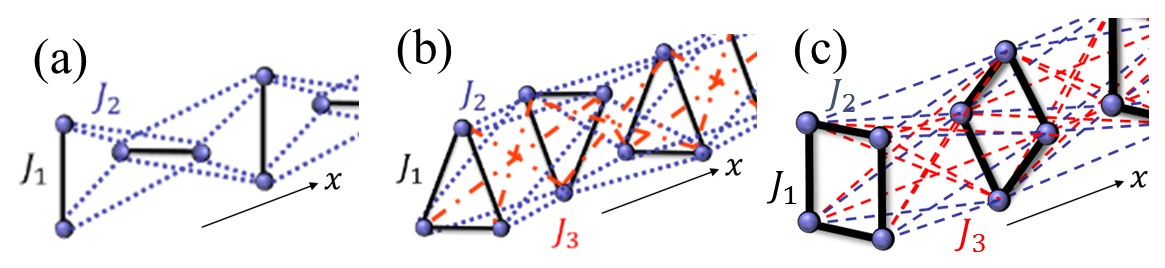}
\caption{(Color online) Multimer chain of (a) dimer ($N=2$), (b) trimer ($N=3$), and (c) tetramer ($N=4$)
with next-nearest neighbor interaction $J_2$ and  third-nearest neighbor interaction $J_3$.
The Hamiltonian, \refeq{eq:H}, is the case $J'=J_2=J_3$.
}
\label{fig:m}
\end{figure}
Schematic figures of the Hamiltonian, \refeq{eq:H},
for $N = 2, 3$, and $4$ are 
shown in Fig.~\ref{fig:m}.
The second term in \refeq{eq:H},
the inter-multimer interaction $J'$ term,
has $N^2$ links. In the $N=4$ case,
next-nearest neighbor interactions and third-nearest neighbor interactions
have the identical strength $J'$ in \refeq{eq:H}.
In this sense, the identical 
inter-multimer interaction is artificial, but it is a key ingredient 
for obtaining exact ground states under the magnetic 
field. In \refeq{eq:H}, the local $S_x$ is 
a good quantum number due to the frustration.\cite{EPJB.15.227}
As a consequence, the model is equivalent to a spin chain model 
with different values of the magnitude of the spin at each site $x$. 


To define the Berry phase 
for the local multimer at a site $x_0$,
we introduce a one-parameter Hamiltonian ${\cal H}(\theta)$ by using the original Hamiltonian ${\cal H}$.
Following Ref.~\citen{EPL.95.20003},
we define
\begin{eqnarray}
  {\cal H}(\theta) = {\cal H} + U(\theta) h_{x_0} U^\dagger(\theta) - h_{x_0}
  \label{eq:H_theta}
  ,
\end{eqnarray}
where $U(\theta)$ is a local gauge twist at the site $x_0$ defined as 
$\displaystyle U(\theta)=\prod_{y=1}^N \e^{\i \phi_y(\theta) (s^z_{x_0,y} - 1/2)}$,
with
\begin{eqnarray}
  \phi_y(\theta)=
  \left\{
  \begin{array}{ll}
    {2\theta (y-1)  \over N}, & 0\leq \theta <\pi
    \\
    {2(2\pi - \theta) (y-1)  \over N} + 2(\theta-\pi) \delta_{y,N}, &\pi \leq \theta \leq 2\pi
    .
  \end{array}
\right.
\nonumber
\end{eqnarray}
Here we choose the spin-twist axis as the $z$-axis 
along the external magnetic field $H$.
For $N=2$,  one can obtain $U(\theta)=\e^{\i \theta (s^z_{x_0,2} - 1/2)}$
and $U(\theta)h_{x_0}U^\dagger(\theta)
=J({\e^{-\i \theta}s_{x_0,1}^+ s_{x_0,2}^- +\e^{\i \theta} s_{x_0,1}^- s_{x_0,2}^+})/2+J s_{x_0,1}^z s_{x_0,2}^z$,
which corresponds to the usual link twist for the $Z_2$ Berry phase\cite{JPSJ.75.123601}.
For general $N$,  
the local gauge twist $U(\theta)$ is equivalent to that in Ref.~\citen{EPL.95.20003}.
A periodicity ${\cal H}(0)={\cal H}(2\pi)={\cal H}$ is easily obtained from the relation $U(0)=U(2\pi)=1$.

For any one-parameter Hamiltonian ${\cal H}(\theta)$ with periodicity ${\cal H}(0)={\cal H}(2\pi)$,
the Berry phase $\gamma$ can be defined in modulo $2\pi$ for $D$-fold states under an energy gap:
\begin{math}
  \gamma = {1\over \i} \oint \mbox{tr} A 
\end{math}
with the $D$-dimensional non-Abelian Berry connection
\begin{math}
  \displaystyle
  (A)_{ij} = 
  \langle \psi_i(\theta) | \d{} \psi_j(\theta)\rangle
  =\bra{\psi_i(\theta)} {\d{} \over \d{}\theta} \ket{\psi_j(\theta)} \d{}\theta
\end{math}
for $D$-fold nearly degenerated ground states
$\ket{\psi_1(\theta)},\ldots ,\ket{\psi_D(\theta)}$.
The Berry phase $\gamma$ is numerically calculated\cite{PRB.47.1651} as
\begin{math}
  \gamma = -\sum_{k=1}^K \arg\det C_k
\end{math}
by using 
discretized integration variables $\theta_k = {2\pi k\over K} +\theta_0$
mod $2\pi$ and a $D\times D$ matrix
\begin{math}
  \left( C_k \right)_{ij} =  \braket{\psi_i(\theta_k)} {\psi_j(\theta_{k+1})}.
\end{math}
Numerical convergence of $\gamma$ is achieved for small $K$
when the Berry phase $\gamma$ is quantized\cite{JPSJ.76.113601}.

For the model [\refeq{eq:H}], quantization of the Berry phase is due to
$y$-direction translational symmetry, that is,
${2\pi \over N}$-rotational symmetry of all $N$-mers.
As shown in Ref.~\citen{EPL.95.20003},
$N \gamma = 0 \mod 2\pi$ is satisfied. Thus, 
the $Z_N$ Berry phase of an $N$-mer, $\gamma_N$, is quantized as 
$\gamma_N=0,{2\pi\over N},{4\pi\over N},\ldots {2(N-1) \pi\over N}$.


Before calculating the $Z_N$ Berry phases of $N$-mer chains,
let us consider a strong multimer coupling limit for the general $N$ case.
In the decoupled limit ($J'=0$), the one-parameter Hamiltonian
satisfies ${\cal H}(\theta) = U(\theta){\cal H}U^\dagger(\theta)$ because of $U(\theta) S^z_{x_0} U^\dagger(\theta)=S^z_{x_0}$.
Then, one obtains the ground states $\ket{\psi_i(\theta)}=U(\theta)\ket{\psi_i(0)}$
and the Berry phase $\displaystyle \gamma_N=2\pi \sum_{i=1}^D \bra{\psi_i(0)}(s^z_{x_0,N}-1/2)\ket{\psi_i(0)}$.
Using the $x$- and $y$-direction translational symmetries of 
the 
$D$-fold degenerated ground states\cite{PRL.78.1984}, 
the Berry phase can be written in terms of the magnetization 
$\displaystyle \langle m\rangle={2 \sum_x \langle  S^z_x \rangle\over LN}$ as 
\begin{eqnarray}
 \gamma_N=\pi (\langle m\rangle-1) D
 .
 \label{eq:gN}
\end{eqnarray}
This corresponds to the fact that $\gamma_N$ depends on the particle number in the decoupled limit of tight-binding models\cite{EPL.95.20003}.
In general, the inter-multimer coupling $J'$ makes $\gamma_N$ non-trivial.

To characterize the Haldane phase in the large $J'$ region,
one can define another quantized $Z_N$ Berry phase $\widetilde{\gamma}_N$ 
by using another one-parameter Hamiltonian
\begin{eqnarray}
  \nonumber
  \widetilde{{\cal H}}(\theta) &=& U^\dagger (\theta) {\cal H}(\theta) U(\theta)
  \\ 
  &=&{\cal H} + U^\dagger(\theta) \tilde{h}_{x_0} U(\theta) - \tilde{h}_{x_0}
  ,
  \label{eq:tildeH}
\end{eqnarray}
where the local inter-multimer Hamiltonian $\tilde{h}_{x_0}$ is defined as
\begin{math}
  \tilde{h}_{x_0}= J'\vec{S}_{x_0-1}\cdot \vec{S}_{x_0}
  +J'\vec{S}_{x_0}\cdot \vec{S}_{x_0+1}
\end{math}
.
In the decoupled limit $J'=0$,
$\widetilde{\gamma}_N$ is zero 
because $\widetilde{{\cal H}}(\theta)$ does not depend on $\theta$. 
Because the Berry phase does not change if the gap-closing does not occur,
the complementary formula \cite{PRB.78.054431} 
\begin{math}
\gamma_N = \widetilde{\gamma}_N + \pi (\langle m\rangle-1) D  
\end{math}
is valid in the small $J'/J$ region.
Whereas $\gamma_N$ is defined by the local gauge twist of the local Hamiltonian $h_{x_0}$ in \refeq{eq:H_theta} to characterize the local multimer,
the complementary Berry phase $\widetilde{\gamma}_N$ is defined by
the local gauge twist of the local inter-multimer Hamiltonian $\tilde{h}_{x_0}$ in \refeq{eq:tildeH} to characterize a local quantum object on the inter-multimer coupling.
It is natural to expect 
$\gamma_N=0$ and $\widetilde{\gamma}_N\neq 0$ in the large $J'/J$ region.
This naive expectation turns out to be true in the Haldane phases,
as shown in the following results.

\begin{figure}
\includegraphics[width=0.5\textwidth]{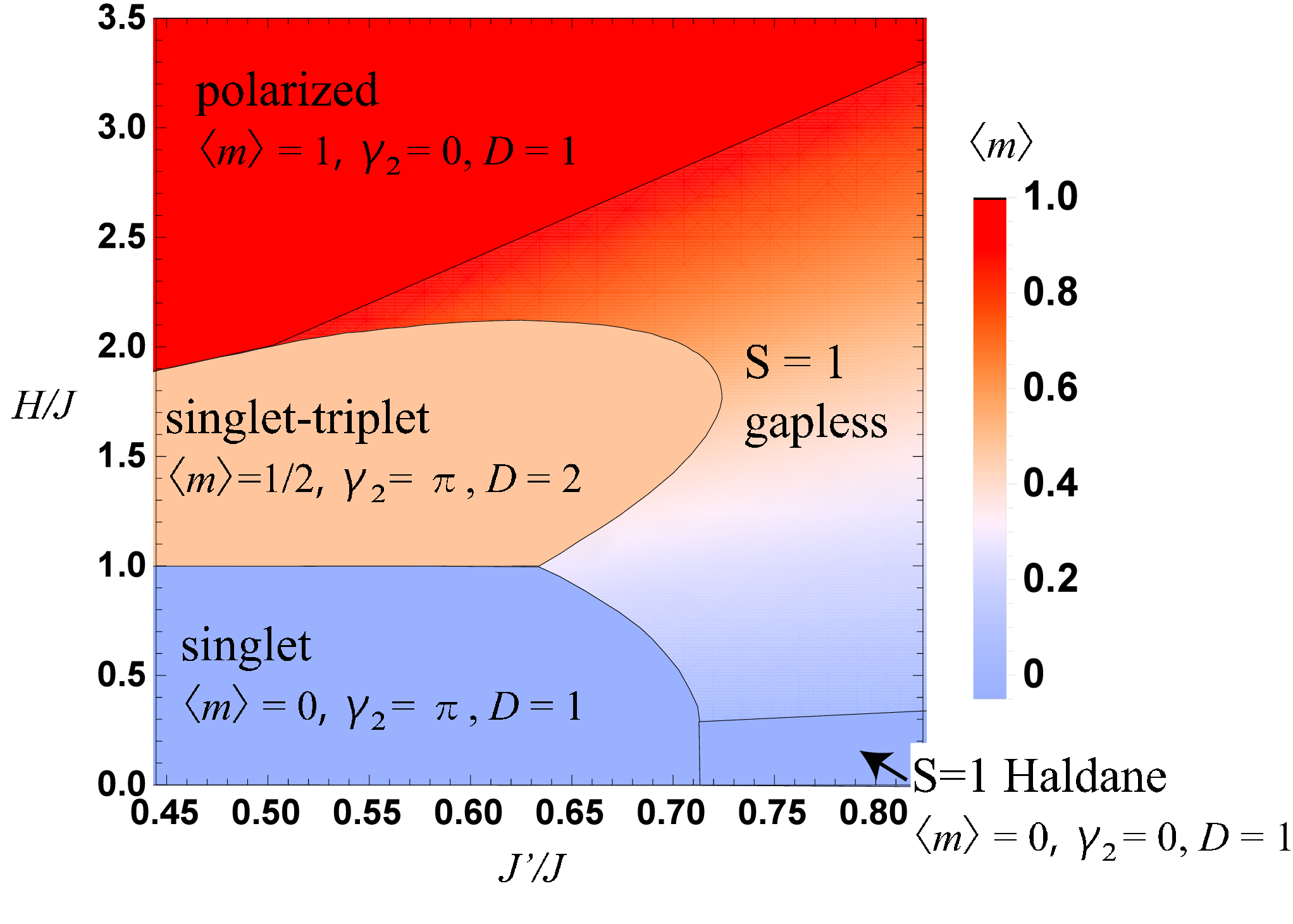}
\caption{(Color online) Phase diagram and $Z_2$ Berry phases for dimer chain ($N=2$).
Phase boundaries\cite{EPJB.15.227} and magnetization curve\cite{PTPS.145.119} are reproduced from each reference.
The complement $Z_2$ Berry phase $\widetilde{\gamma}_2$
is $\widetilde{\gamma}_2=\pi$ in the $S=1$ Haldane phase.
}
\label{fig:phaseN2}
\end{figure}
For the dimer case ($N=2$), the phase diagram and its magnetization plateaux have been studied\cite{EPJB.15.227}.
By using the phase boundaries and magnetization curves in Ref.~\citen{EPJB.15.227},
the phase diagram is reproduced in Fig.~\ref{fig:phaseN2}.
In the small $J'/J$ region, 
the ground state of each magnetization plateaux is exactly written as the 
direct product state of the dimer singlet and triplet.
On the other hand, in the large $J'/J$ region, 
the ground state of the model is equivalent to that of the $S=1$ Heisenberg chain.

The phases are classified with $Z_2$ Berry phases $\gamma_2=0$ or $\pi$ of the $D$-fold degenerated ground states except for the gapless phase, as shown in Fig.~\ref{fig:phaseN2}.
It should be noted that $\gamma_2$ in each plateau phase requires numerical calculation because the ground states of ${\cal H}(\theta)$ are not simple direct product states 
when $\theta \neq 0$.  
$\gamma_2$ in the dimer-singlet phase ($\langle m \rangle = 0$) and
singlet-triplet phase ($\langle m \rangle = 1/2$) are equivalent to those in the decoupled limit ($J'=0$),
{\it i.e.}, $\gamma_2=\pi (\langle m\rangle-1) D$.
When $J'=0$,
the local states for each total $S_z$ sector and its local Berry phase $\gamma_2$ are easily obtained:
the spin singlet $\ket{s}=(\ket{\uparrow\downarrow}-\ket{\downarrow\uparrow})/\sqrt{2}$ with $\gamma_2=\pi$
for $\langle m \rangle=0$, 
and
the fully polarized $\ket{t}=\ket{\uparrow\uparrow}$ with $\gamma_2=0$
for $\langle m \rangle=1$.
Even for the finite $J'$ case in Fig.~\ref{fig:phaseN2},
the exact ground state of the singlet phase is 
$\ket{ss\cdots}=\ket{s}\ket{s}\cdots$ with $\gamma_2 =\pi$
and that of the polarized phase is $\ket{tt\cdots}$ with $\gamma_2 = 0$.
In short, $\gamma_2$ detects the singlet on a dimer.
In the singlet-triplet phase,
the exact ground states are 
doubly degenerated; 
$\ket{stst\cdots}$
and 
$\ket{tsts\cdots}$.
The Berry phase of the doubly degenerated ground states ($D=2$)
can be understood as the summation of 
$\pi$ for $\ket{s}$ and 
$0$ for $\ket{t}$.
Then, we obtain $\gamma_2=\pi+0=\pi$ for the singlet-triplet phase.
Finally, in the $S=1$ Haldane phase,
the ground state is not exact but the valence bond solid (VBS) type.
Because the spin singlets of the VBS state exist on the inter-dimer links,
it is natural that the ground state of the $S=1$ Haldane phase has $\gamma_2=0$ 
and $\widetilde{\gamma}_2=\pi$.

\begin{figure}
\includegraphics[width=0.5\textwidth]{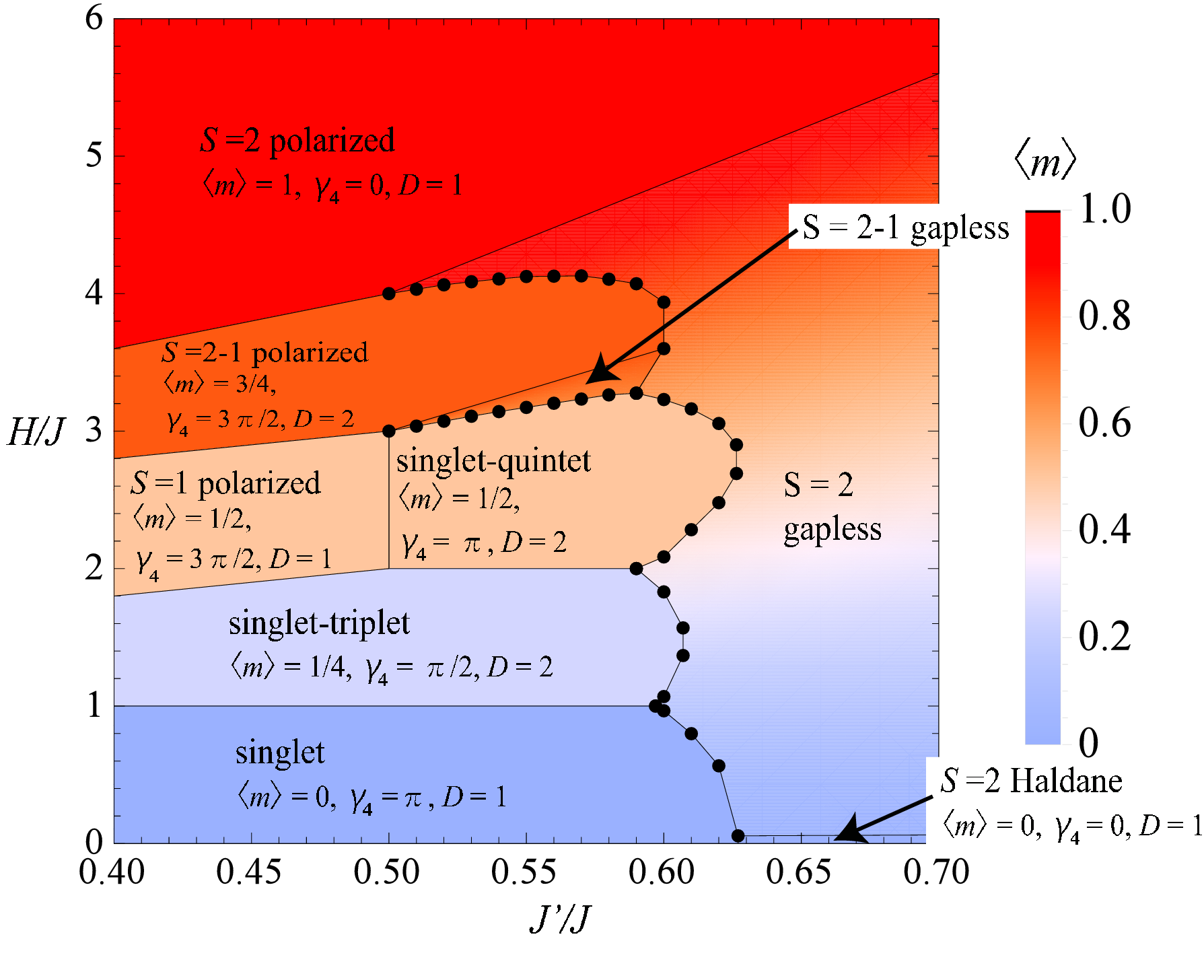}
\caption{(Color online) Phase diagram for the tetramer chain ($N=4$) in the exact case ($J_3=J_2$).
Each magnetization plateau is classified by its magnetization $\langle m \rangle$ and $Z_4$ Berry phase $\gamma_4$ for $D$-fold degenerated ground state.
Filled circles on phase boundaries are obtained by the Lanczos method.
The transition point to the Haldane phase is determined by its energy\cite{PRL.87.047203}.
The complement $Z_4$ Berry phase $\widetilde{\gamma}_4$
is $\widetilde{\gamma}_4=\pi$ in the $S=2$ Haldane phase.
}
\label{fig:phaseN4}
\end{figure}

As shown in Fig.~\ref{fig:phaseN4},
the phase diagram of the tetramer chain ($N=4$) can be classified by
the $Z_4$ Berry phase: $\gamma_4=0, {\pi \over 2}, \pi$, and$ {3\pi \over 2}$.
The phase boundaries, which are equivalent to those for the $S=1$ highly frustrated ladder\cite{PRB.74.024421}, were determined by comparing the energies between 
the exact direct product states, $S=2$ chain with $L=12$, and $S=1$-$2$ chain with $L=12$.
The exact direct product states are described as the products of 
singlet $\ket{s}={\ket{\uparrow\uparrow\downarrow\downarrow} - 2\ket{\uparrow\downarrow\uparrow\downarrow}
+\ket{\uparrow\downarrow\downarrow\uparrow}
+\ket{\downarrow\uparrow\uparrow\downarrow}
-2\ket{\downarrow\uparrow\downarrow\uparrow}
+\ket{\downarrow\downarrow\uparrow\uparrow}
\over 2\sqrt{3}}$ ($\langle h_x \rangle = -2J$),
triplet $\ket{t}={\ket{\uparrow\uparrow\uparrow\downarrow}-\ket{\uparrow\uparrow\downarrow\uparrow}+\ket{\uparrow\downarrow\uparrow\uparrow}-\ket{\downarrow\uparrow\uparrow\uparrow} \over 2}$ ($\langle h_x \rangle = -J$),
and quintet $\ket{q}=\ket{\uparrow\uparrow\uparrow\uparrow}$ ($\langle h_x \rangle = J$) states on the tetramer.
The energies of the $S=2$ and $S=1$-$2$ chains were calculated by the Lanczos method.
Note that the transition point to the $S=2$ Haldane phase is determined from 
the $S=2$ Haldane state energy obtained by the Monte Carlo method in Ref.~\citen{PRL.87.047203}.
The Berry phases in Fig.~\ref{fig:phaseN4} are numerically integrated from a few low-energy states obtained by exact diagonalization up to 16 spins ($L\leq 4$ and $N=4$).
Using the fact that the Berry phases are identical unless energy-level crossing occurs,
one can determine Berry phases continuously from the decoupled limit to the small $J'/J$ region
by just calculating the low energies by the Lanczos method.

As a result, the decoupled limit formula, $\gamma_4=\pi (\langle m\rangle-1) D$, is satisfied
for all plateau
phases except for the $S=2$ Haldane phase.
For example, the singlet-quintet phase has doubly degenerated ground states:
$\ket{sqsq\cdots}$ and $\ket{qsqs\cdots}$.
After the summation of the two Berry phases at site $x_0$, 
$\pi$ for $\ket{s}$ and $0$ for $\ket{q}$,
we obtain $\gamma_4=\pi$ for the doubly degenerated states in the singlet-quintet phase.
In the other five phases except for the Haldane phase,
the ground states of the untwisted Hamiltonian ${\cal H}$ and the Berry phase $\gamma_4$ of the corresponding plateau phase
are
$\ket{ss\cdots}$ and $\gamma_4=\pi$ for the singlet phase,
and
$\ket{st\cdots}$ and $\ket{ts\cdots}$, $\gamma_4=\pi/2$ for the singlet-triplet phase,
$\ket{tt\cdots}$ and $\gamma_4=3\pi/2$ for the $S=1$ polarized phase,
$\ket{tq\cdots}$ and $\ket{qt\cdots}$, $\gamma_4=3\pi/2$ for the $S=2-1$ polarized phase,
$\ket{qq\cdots}$ and $\gamma_4=0$ for the $S=2$ polarized phase.
It should be emphasized that 
this simple understanding is due to the simplicity of 
the present Hamiltonian with fully connected inter-multimer interaction\cite{EPJB.15.227}.
Finally, in the $S=2$ Haldane phase,
the ground state is not exact but the VBS type.
The Berry phases $\gamma_4=0$ and $\widetilde{\gamma}_4=\pi$ in the $S=2$ Haldane phase
can be understood in the same way as $\gamma_2=0$ and $\widetilde{\gamma}_{2}=\pi$ in the $S=1$ Haldane phase of Fig.~\ref{fig:phaseN2};
that is, the VBS state is disentangled on local-multimer sites and entangled on inter-multimer links.
It will be an interesting problem if the $S=1$ Haldane phase appears between the singlet phase and
 the $S=2$ Haldane phase.
However, a comparison with numerical energies\cite{PRL.87.047203} revealed that 
the $S=1$ Haldane phase does not appear along the $H=0$ line of the phase diagram
for the present model, nor for the model of Ref.~\citen{EPJB.15.227}.


\begin{figure}
  \begin{center}
\includegraphics[width=0.4\textwidth]{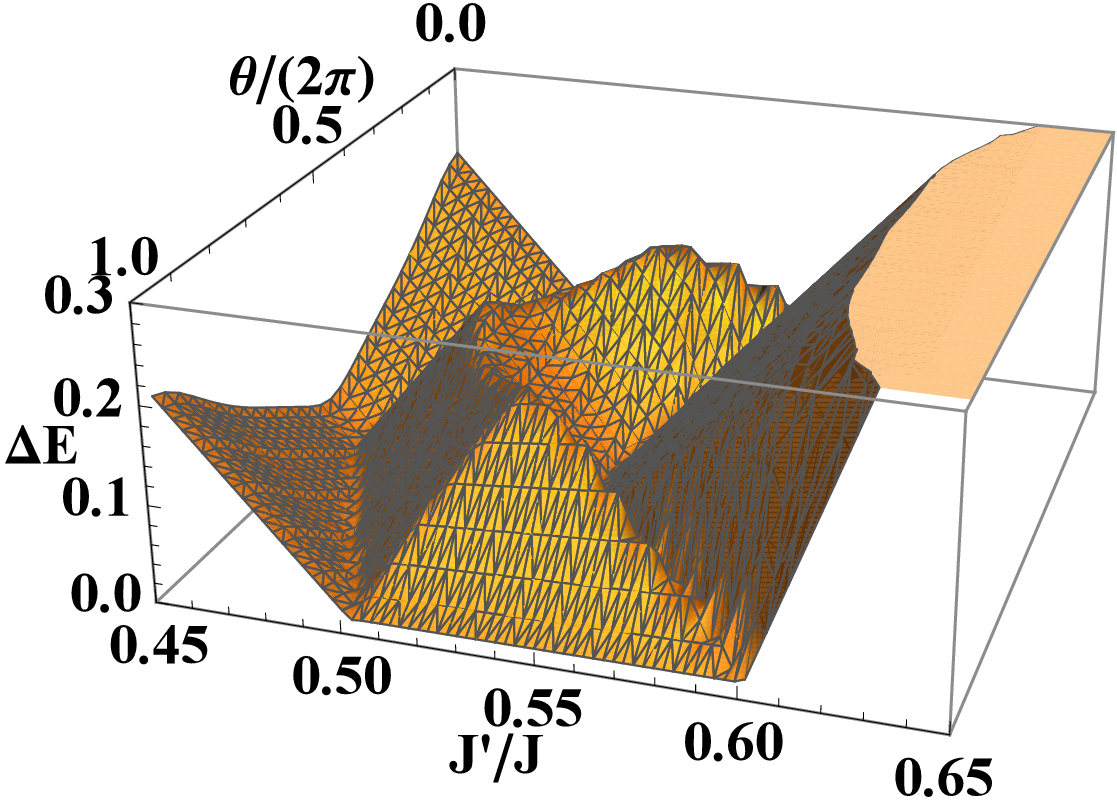}
  \end{center}
\caption{(Color online) Excitation gap $\varDelta E$ as a function of twist angle $\theta$ and
  $J'/J$ in the tetramer ($N=4$) chain at $\langle m \rangle=1/2$.  }
\label{fig:gapN4}
\end{figure}
An advantage of the Berry phase is the ability to identify a phase transition even if gap-closing does not occur due to the finite-size gap for small system size,
as demonstrated in bond-alternating models for $Z_N$ Berry phases\cite{EPL.95.20003}.
In the present model,
because the phase transition between the plateau phases is a simple first-order transition\cite{EPJB.15.227},
gap-closing at a transition point exists for small system size
and could persist for $\theta>0$ due to the symmetry difference of the ground states\cite{JPCONF.320.012019}.
To check a detail of the gap-closing line near a phase boundary,
the excitation gap $\varDelta E$ is calculated by exact diagonalization for $\langle m \rangle=1/2$ sector of 16 spins ($L=4$ and $N=4$).
In Fig.~\ref{fig:gapN4},
the excitation gap $\varDelta E$ is shown as a function of 
$\theta$ and $J'/J$
around the transition from the $S=1$ polarized phase through the singlet-quintet phase to the gapless phase.
Here, 
the double degeneracy of $\ket{sq\cdots}$ and $\ket{qs\cdots}$ in the singlet-quintet phase
corresponds to
$\varDelta E=0$ at $\theta=0$ around $0.5 \le J'/J \le 0.6$.
For $\theta>0$, this double degeneracy is lifted 
because
the local quintet $\ket{q}$ at the site $x_0$ cannot be affected by $\theta$
while the local singlet stabilizes.
This kind of gap-opening can be observed by the alternating strength of $J$.
With the use of this gap-opening, we can define each Berry phase for the degenerated states,
and we obtain $\gamma_4=\pi$ for the local singlet and $\gamma_4=0$ for the local quintet.


Finally, let us comment on the other $N$ cases.
An analysis of the even $N$ case is similar to that of the $N=2$ and 4 cases.
At least in the weak coupling region, exact ground states can be obtained by the product state of the
decoupled limits ($J' = 0$), and the fractionally quantized Berry phase can be defined by
$\gamma_N=\pi (\langle m\rangle-1) D$.
When $N$ is odd, the plateaux for $\langle m \rangle = 0$ may vanish, because
the model can be mapped to a 1D Heisenberg model of half integer spins $S = 1/2, 3/2, \cdots, N/2$. 
However, there is room to obtain plateau phases for $\langle m \rangle \neq 0$.
In fact, $\langle m \rangle = 1/3$ and $2/3$ plateaux for $N=3$ 
are stabilized in Ref.~\citen{EPJB.15.227}.
In most of the plateau phases for $N = 3$,  
there is macroscopic degeneracy and/or residual entropy due to degeneracy of the local multimer.
However, $\gamma_N=\pi (\langle m\rangle-1) D$ seems valid even in such plateau phases.
Note that the degeneracy $D$ is macroscopically large in such plateaux, and
$\gamma_N$ depends on the system size $L$.


In conclusion, 
fractionally quantized $Z_N$ Berry phases 
have classified magnetization plateaux
in the spin-1/2 Heisenberg multimer ($N$-mer) chain
as an extension of the orthogonal dimer ($N=2$) chain.
In addition to a transition from a zero Berry phase to a non-zero Berry phase,
which is the usual case in the previous studies on the quantized Berry phase,
the results in this letter show that one phase diagram can be classified by all the $N$ types of Berry phases $\gamma_N=0, {2\pi \over N}, {4\pi \over N},\ldots {2(N-1)\pi \over N}$
that are defined for identification of the local multimer.
Moreover, we have found the complementary Berry phase $\widetilde{\gamma}_N$,
which becomes $\widetilde{\gamma}_N=\pi $ in the Haldane phases.
This demonstration is evidence that
the $Z_N$ Berry phase has the potential to characterize various gapped phases.

Because the present model is a simple model to understand the physics of the magnetization plateaux and jumps,
the simple interpretation of $\gamma_N$ has been obtained from the decoupled limit formula, \refeq{eq:gN},
except for the Haldane phase.
This decoupled limit formula seems valid 
even when the ground state cannot be represented by the direct product states,
unless energy-gap closing occurs.
We checked it with an exact diagonalization for 
the model in Fig.~\ref{fig:m} (c) with 
$J_2 \neq J_3$.
One of the future problems is phase determination for
the general case, {\it e.g.}, $J_2 \neq J_3$ in Figs.~\ref{fig:m}(b) and (c). 

Another future problem is to characterize the general $S > 2$ Haldane phase
with the complementary Berry phase $\widetilde{\gamma}_{2S}$,
which has successfully characterized the $S=2$ Haldane phase with the non-trivial Berry phase $\widetilde{\gamma}_4=\pi$.
It should be emphasized that the conventional $Z_2$ Berry phase defined by the link twist becomes zero in the even $S$ Haldane phase,
reflecting the even number of spin singlets existing on a link.
From this viewpoint, we have found a new topological invariant for the $S=2$ Haldane phase.
These Berry phases, ${\gamma}_{N}$ and  $\widetilde{\gamma}_{N}$, can be applied 
to a wide class of models because
the quantization of the $Z_N$ Berry phases holds for any Hamiltonian with rotational symmetry of the local $N$-mer.

\acknowledgment 
The authors thank H. Ueda and T. Oka for useful discussions.
The work is partially supported by JSPS KAKENHI
Grand Number JP17H02926.
The computation was partially carried out using the computer facilities at Research Institute for Information Technology, Kyushu University.

\begin{thebibliography}{10}

\bibitem{PRB.45.304}
T. Kennedy and H. Tasaki:  Phys. Rev. B {\bf 45} (1992) 304.
\bibitem{PRB.40.4709}
M. den Nijs and K. Rommelse:  Phys. Rev. B {\bf 40} (1989) 4709.
\bibitem{PRL.89.077204}
M. Nakamura and S. Todo:  Phys. Rev. Lett. {\bf 89} (2002) 077204.
\bibitem{PRB.81.064439}
F. Pollmann, A.~M. Turner, E. Berg, and M. Oshikawa:  Phys. Rev. B {\bf 81}
  (2010) 064439.
\bibitem{JPSJ.75.123601}
Y. Hatsugai:  J. Phys. Soc. Jpn. {\bf 75} (2006) 123601.
\bibitem{PRL.49.405}
D.~J. Thouless, M. Kohmoto, M.~P. Nightingale, , and M. den Nijs:  Phys. Rev.
  Lett. {\bf 49} (1982) 405.
\bibitem{PRB.94.205112}
N. Chepiga, I. Affleck, and F. Mila:  Phys. Rev. B {\bf 94} (2016) 205112.
\bibitem{EPL.95.20003}
Y. Hatsugai and I. Maruyama:  Europhys. Lett. {\bf 95} (2011) 20003.
\bibitem{AX.1508.00960}
Y. Motoyama and S. Todo:  ArXiv: 1508.00960 (2015) .
\bibitem{AX.1709.01546}
T. Kariyado, T. Morimoto, and Y. Hatsugai:  Phys. Rev. Lett. {\bf 120} (2018)
  247202.
\bibitem{AX.1806.10767}
T. Kawarabayashi, K. Ishii, and Y. Hatsugai:  ArXiv: 1806.10767 (2018) .
\bibitem{EPJB.15.227}
A. Honecker, F. Mila, and M. Troyer:  Eur. Phys. J. B {\bf 15} (2000) 227.
\bibitem{PRB.43.8644}
M.~P. Gelfand:  Phys. Rev. B {\bf 43} (1991) 8644.
\bibitem{PB.108.1069}
B.~S. Shastry and B. Sutherland:  Physica B {\bf 108} (1981) 1069.
\bibitem{JPCM.15.327}
S. Miyahara and K. Ueda:  J. Phys: cond. Matt. {\bf 15} (2003) 327.
\bibitem{JPCONF.320.012019}
I. Maruyama, S. Tanaya, M. Arikawa, and Y. Hatsugai:  J. Phys.: Conf. Ser. {\bf
  320} (2011) 012019.
\bibitem{PRB.47.1651}
R.~D. King-Smith and D. Vanderbilt:  Phys. Rev. B {\bf 47} (1993) 1651.
\bibitem{JPSJ.76.113601}
I. Maruyama and Y. Hatsugai:  J. Phys. Soc. Jpn. {\bf 76} (2007) 113601.
\bibitem{PRL.78.1984}
M. Oshikawa, M. Yamanaka, and I. Affleck:  Phys. Rev. Lett. {\bf 78} (1997)
  1984.
\bibitem{PRB.78.054431}
T. Hirano, H. Katsura, and Y. Hatsugai:  Phys. Rev. B {\bf 78} (2008) 054431.
\bibitem{PTPS.145.119}
K. Okunishi:  Prog. Theor. Phys. Suppl. {\bf 145} (2002) 119.
\bibitem{PRL.87.047203}
S. Todo and K. Kato:  Phys. Rev. Lett. {\bf 87} (2001) 047203.
\bibitem{PRB.74.024421}
V.~R. Chandra and N. Surendran:  Phys. Rev. B {\bf 74} (2006) 024421.
\end{thebibliography}

\end{document}